\documentclass{article}

\usepackage{PRIMEarxiv}
\usepackage{longtable}
\usepackage[utf8]{inputenc} 
\usepackage[T1]{fontenc}    
\usepackage{hyperref}       
\usepackage{url}            
\usepackage{booktabs}       
\usepackage{amsfonts}       
\usepackage{nicefrac}       
\usepackage{microtype}      
\usepackage{lipsum}
\usepackage{fancyhdr}       
\usepackage{graphicx}       
\graphicspath{{media/}}     
\usepackage{txfonts}
\usepackage{xcolor}

\title{The Distribution Function of the Average Iron Charge State at 1 AU:
From a Bimodal Wind to ICME Identification
\thanks{\textit{\underline{Citation}}: 
\textbf{Larrodera, C. and C. Cid (2020b). “The Distribution Function of the Average Iron Charge
State at 1 AU: From a Bimodal Wind to ICME Identification”. In: Solar Physics 295.11,
156, p. 156. DOI: 10.1007/s11207-020-01727-8.}} 
}

\author{
  C. Larrodera \\
  University of Alcalá \\
  Alcalá de Henares\\
  \texttt{carlos.larrodera@edu.uah.es} \\
   \And
  C. Cid \\
  University of Alcalá \\
  Alcalá de Henares\\
  \texttt{consuelo.cid@uah.es} \\
}

\begin{document}
\maketitle

\begin{abstract}
We aim to investigate the distribution function of the iron charge state, at 1AU to check if it corresponds to a bimodal wind. We use data from Solar Wind Ion Composition Spectrometer (SWICS) instrument on board the Advanced Composition Explorer (ACE) spacecraft along 20 years. We propose the bi-Gaussian function as the probability distribution function that fits the $\langle Q_{Fe}\rangle$ distribution. We study the evolution of the parameters of the bimodal distribution with the solar cycle. We compare the outliers of the sample with the existing catalogs of Interplanetary Coronal Mass Ejections (ICMEs) and identify new ICMEs.
The $\langle Q_{Fe}\rangle$ at 1 AU shows a bimodal distribution related to the solar cycle. Our results confirm that $\langle Q_{Fe}\rangle> 12 $ is a trustworthy proxy for ICME identification and a reliable signature in the ICME boundary definition.
\end{abstract}


\section{Introduction}
The existence of solar wind was first confirmed by in situ spacecrafts in the 1960s \cite{Gringauz_1960_Solar_Wind,Gringauz_1961_Solar_wind,Gringauz_1967_Solar_wind}.
\cite{Neugebauer_1966_Solar_Wind_Mariner2} showed for the first time a bimodal wind, based on Mariner 2 observations of recurring streams of high speed plasma.
From mid 1970s on, new space missions with new instruments have studied the solar wind using different physical magnitudes like proton speed, proton temperature or interplanetary magnetic field among others \cite{Hundhausen_1972_Solar_Wind_review,Rossi_1991_Interplanetary_plasma}.

Nowadays, the solar wind is commonly classified into slow and fast (which form the bulk of the wind) and transient events coming from  coronal mass ejections, where material from the solar atmosphere is thrown into the interplanetary space \cite{2006_Schwenn_solar_wind_review, Viall2020}.
Each type of bulk wind has a different source. The fast solar wind is considered  to come from the open magnetic field that emerges from the Sun through the coronal holes \cite{Banaszkiewicz_1997_Fast_SW,Schwenn_2006_Solar_wind}. The source of the slow solar wind is not clear, and many suggestions have been made, e.g. plasma released by reconnection between open and closed magnetic field lines \cite{2005_Lionello_slow_wind} or flow emerging from small equatorial coronal holes \cite{Bale_2019_Solar_wind_CH}.
Nevertheless, \cite{Neugebauer_2002_Solar_Wind}, from a study of the sources of the solar wind during the maximum of Solar Cycle 23 (1998-2001), conclude that near the solar maximum the characteristics of the fast solar wind are very different from the fast solar wind during the minimum when the polar coronal holes are dominant because of the contribution of the active regions to the fast stream.
They suggest a hierarchy based on open field regions with large polar coronal holes as the source of the highest speeds, and active regions like other contributors to the fast wind showing a correlation between the fast wind and the solar cycle.

Through the years different distribution functions have been used to model different solar wind  parameters, e.g. a lognormal function  \cite{Burlaga_1979_IMF} or kappa-like \cite{Voros_2015_Solar_Wind_Statistics} for the magnetic field, or a gamma-like function \cite{Li_2016_Solar_wind_Statistical_Analysis} for solar wind speed, among others.
Recently \cite{Larrodera_2020_Bimodal_solar_wind} proposed the bi-Gaussian function, as a characterization of the solar wind at 1 AU. This model has been applied to the distribution function, not only of the proton speed, but also of the proton temperature, magnetic field and density, showing a bimodal distribution of all the four solar wind parameters.

While the solar wind speed or the proton density can dynamically evolve between the Sun and the Earth as a result of the interactions of the different streams coming out from the Sun, no major changes are expected in the composition of the solar wind, which is frozen since its departure from the Sun. Thus, the solar wind composition can be considered a useful physical magnitude to characterize it \cite{Cranmer_2017_Solar_Wind_Review, Schwenn_2006_Solar_wind}.
Indeed, the solar wind composition is used as a signature of fast streams coming from coronal holes (e.g. \cite{Heidrich_2016_Composition}) and
interplanetary coronal mass ejections (ICMEs).
In particular, the average iron charge state ($\langle Q_{Fe}\rangle$), 
has been used to identify ICMEs \cite{Lepri_2001_ICME_QFe, Lepri_2004_ICME_Composition} or to establish their boundaries \cite{Cid_2016_ICME_Boundaries}. 

In this article we study the distribution function of $\langle Q_{Fe}\rangle$ measured by the Solar Wind Ion Composition Spectrometer (SWICS) instrument \cite{Gloeckler_1998_SWICS} on board the Advance Composition Explorer (ACE). Our first aim is to check if the bi-Gaussian function is able to separate each contribution from the slow and fast winds. We also study the relevance of ICMEs in the solar wind sample and how the bimodal approach is related with the solar cycle.
Section \ref{sec:data} describes the data set used in the study. Section \ref{sec:approach} applies the bi-Gaussian approach to the distribution function of $\langle Q_{Fe}\rangle$ at 1 AU, first to the whole data sample and then to a reduced data sample, after removing ICMEs and outliers from the bulk solar wind. Then, in Section \ref{sec:outliers_ICMEs} we analyze the outliers of the sample, comparing to previously identified ICMEs. 
Finally, Section \ref{sec:conclusions} details our conclusions.

\section{Data} \label{sec:data}

This study uses data from SWICS on board ACE. Specifically, we use level 2 average iron charge state data with two-hour resolution from the ACE Science Data Center (\urlstyle{same}{\url{http://www.srl.caltech.edu/ACE/ASC/level2/index.html}}). 

\begin{figure}[ht]
	\includegraphics[width=\columnwidth]{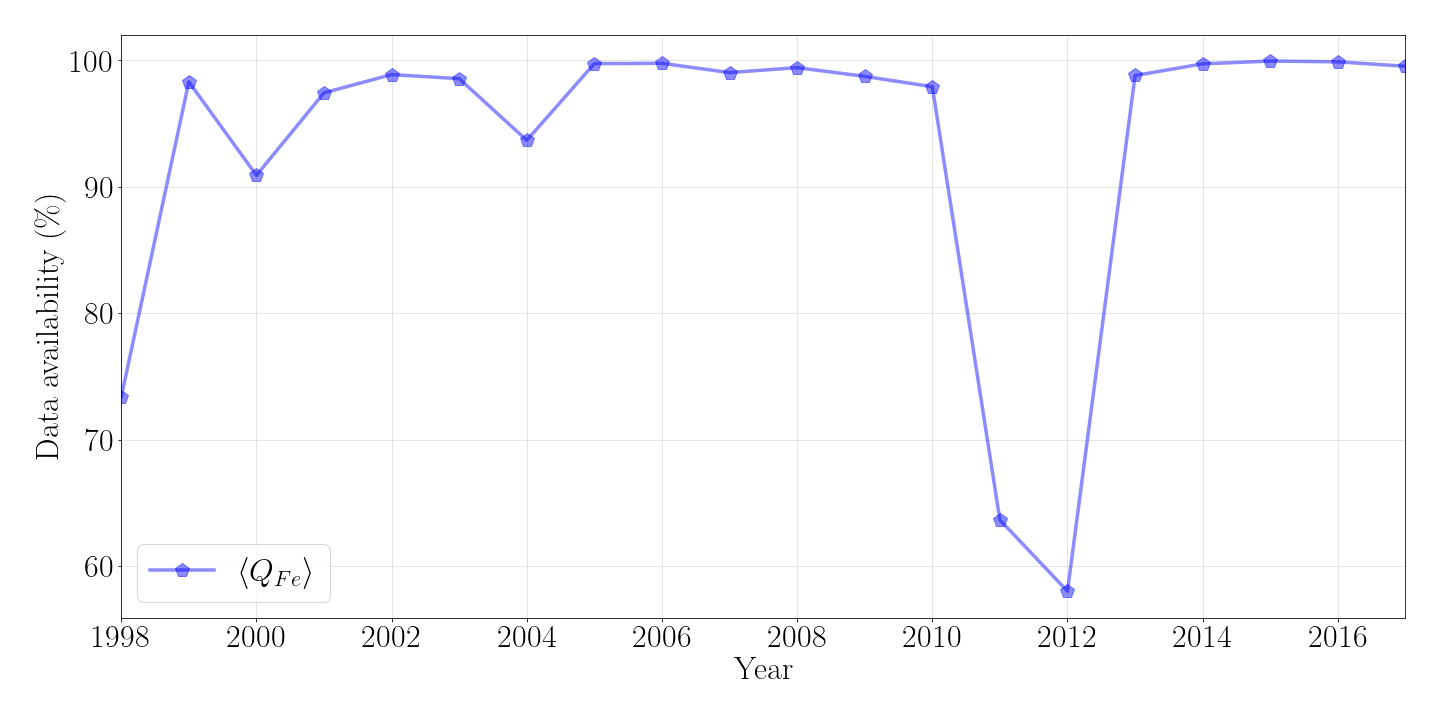}
    \caption{Data availability of $\langle Q_{Fe}\rangle$ measured by SWICS/ACE.}
    \label{fig:data_coverage}
\end{figure}

Figure \ref{fig:data_coverage} shows the data availability during the period under study (1998-2017). The drop between 2010-2013 was due to a radiation and age-induced hardware anomaly which altered the instrument operational state, as informed by the operational SWICS team. Nevertheless the lowest availability value is around 60 \% which is high enough for the analysis. 

Indeed, there are two different SWICS data sets: SWICS 1.1, prior to the 23$^{th}$ August 2011 anomaly, and SWICS 2.0, for the time period after the anomaly. 
According to the Data Release Notes from the instrument team, SWICS 1.1 data provide a reliable ground truth for validation of methods to obtain SWICS 2.0 data. 
Therefore, we have checked the results obtained from the whole SWICS data set against those obtained only using SWICS 1.0 data. Although the data in this sample is reduced in this case, we consider that the results obtained from the whole SWICS data set are robust (and therefore described in this article) if they coincide with those from SWICS 1.0 data set.

\section{A Bi-Gaussian approach for $\langle Q_{Fe}\rangle$} \label{sec:approach}

We propose a bi-Gaussian function as the probability distribution function (PDF) for $\langle Q_{Fe}\rangle$ at 1 AU. This function is defined as the addition of two Gaussian distribution functions where each one represents the contributions of one type of solar wind. A bi-Gaussian function has been previously used to explain the distribution of other solar wind magnitudes \cite{Larrodera_2020_Bimodal_solar_wind}. 

The analytical expression of the bi-Gaussian function in this case is:

\begin{equation}
bG(\langle Q_{Fe}\rangle)=h_{1}\cdot exp\left(\frac{-\left(\langle Q_{Fe}\rangle-p_{1}\right)^{2}}{2w_{1}^{2}}\right)+h_{2}\cdot exp\left(\frac{-\left(\langle Q_{Fe}\rangle-p_{2}\right)^{2}}{2w_{2}^{2}}\right)
\end{equation}
where $h_{1}$, $p_{1}$, $w_{1}$, $h_{2}$, $p_{2}$ and $w_{2}$ are the parameters obtained from the fitting to the data set. The subscripts of these parameters correspond to the first (1) or second (2) Gaussian distribution function, respectively,  representing the two types of wind. $h$ represents the height of the peak of each single Gaussian, $p$ the position of the peak, and $w$ the root-mean-square (RMS) width of each single Gaussian.

\begin{figure}[ht]
	\includegraphics[width=\columnwidth]{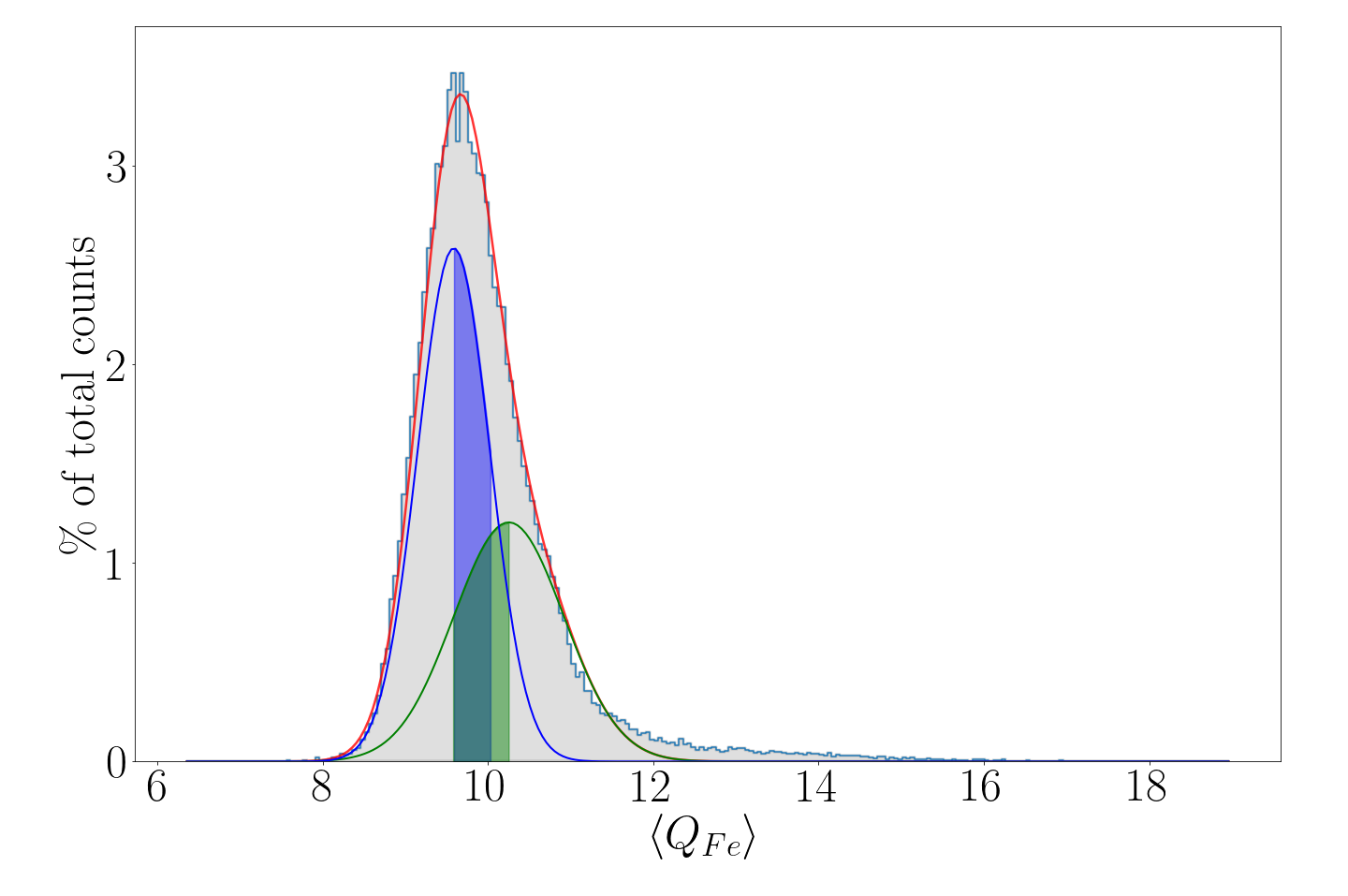}
    \caption{Empirical distribution function of $\langle Q_{Fe}\rangle$ for the whole data set  under study from SWICS/ACE (grey) and the fitting to a bi-Gaussian function (red). Green and blue lines correspond to the single Gaussian curves. See text for explanation on blue and green shadowed areas.}
    \label{fig:qfe_all}
\end{figure}

Figure \ref{fig:qfe_all} shows the normalized distribution of the whole data set of $\langle Q_{Fe}\rangle$ and the fitting to a bi-Gaussian function. Blue and green lines represent the individual Gaussian functions for each type of wind. 
Most of $\langle Q_{Fe}\rangle$ is spread around $p_{1}=9.6$ but there is also another significant contribution around $p_{2}=10.3$.
The Pearson-$\chi^{2}$ value from the fitting is 4.4 with a data availability of 93.3\%, showing that a bimodal distribution can be considered.
Nevertheless, assuming $w_{1}=0.4$ and $w_{2}=0.7$ as the uncertainties of $p_{1}$ and $p_{2}$, respectively, we obtain that the position of both peaks overlaps. This is shown by shadowed regions: the blue region covers the interval ($p_{1}$, $p_{1}+w_{1}$) and the green region the ($p_{2}-w_{2}$, $p_{2}$) one.
The small difference between the position of the center of the peak of each single Gaussian questions the bi-Gaussian approach for $\langle Q_{Fe}\rangle$. Indeed, yearly samples might have missed 
a detailed behaviour which is expected to arise in shorter samples. Thus we study the monthly evolution of $\langle Q_{Fe}\rangle$ distribution function with the aim of solving this concern.

\begin{figure}[ht]
	\includegraphics[width=\columnwidth]{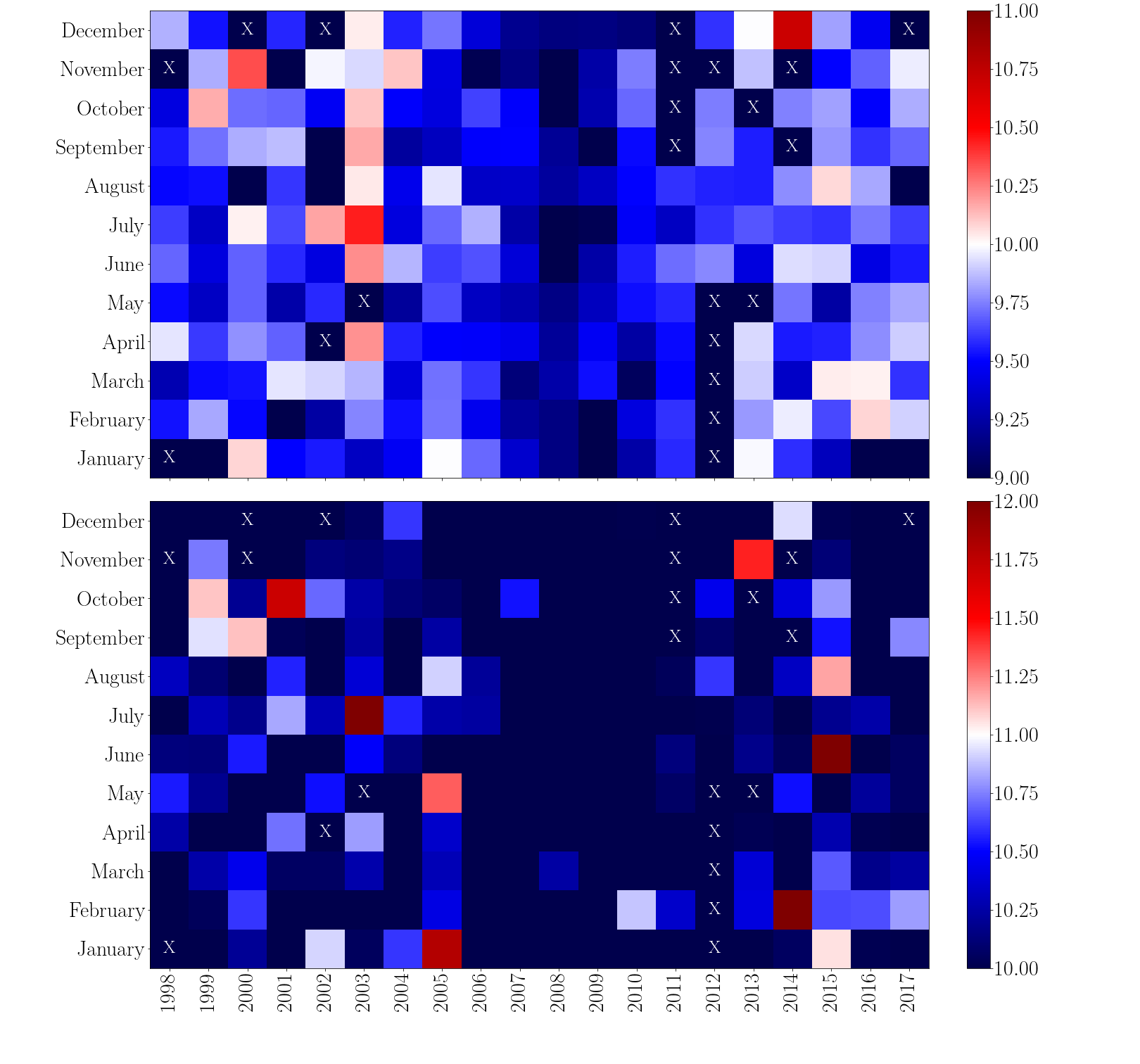}
    \caption{ Positions $p_{1}$ and $p_{2}$ from the bi-Gaussian fitting of the monthly $\langle Q_{Fe}\rangle$ distribution functions. Top (bottom) represents the first (second) peak. The white 'Xs' marks the months where the fitting was not possible.}
    \label{fig:monthly_gauss}
\end{figure}

Figure \ref{fig:monthly_gauss} represents the position of the peak of the first (top) and second (bottom) Gaussian functions for each month along the whole data set. Note that the range of the color scale in both figures is slightly different.
We observe three vertical bands which match the periods of the solar cycle: (1) from the ascending to the declining phase of Solar Cycle 23 (1998-2005), (2) during the minimum (2006-2012), and (3) from the ascending to the declining phase of Solar Cycle 24 (2013-2017). 

\begin{figure}[h]
	\includegraphics[width=0.96\columnwidth]{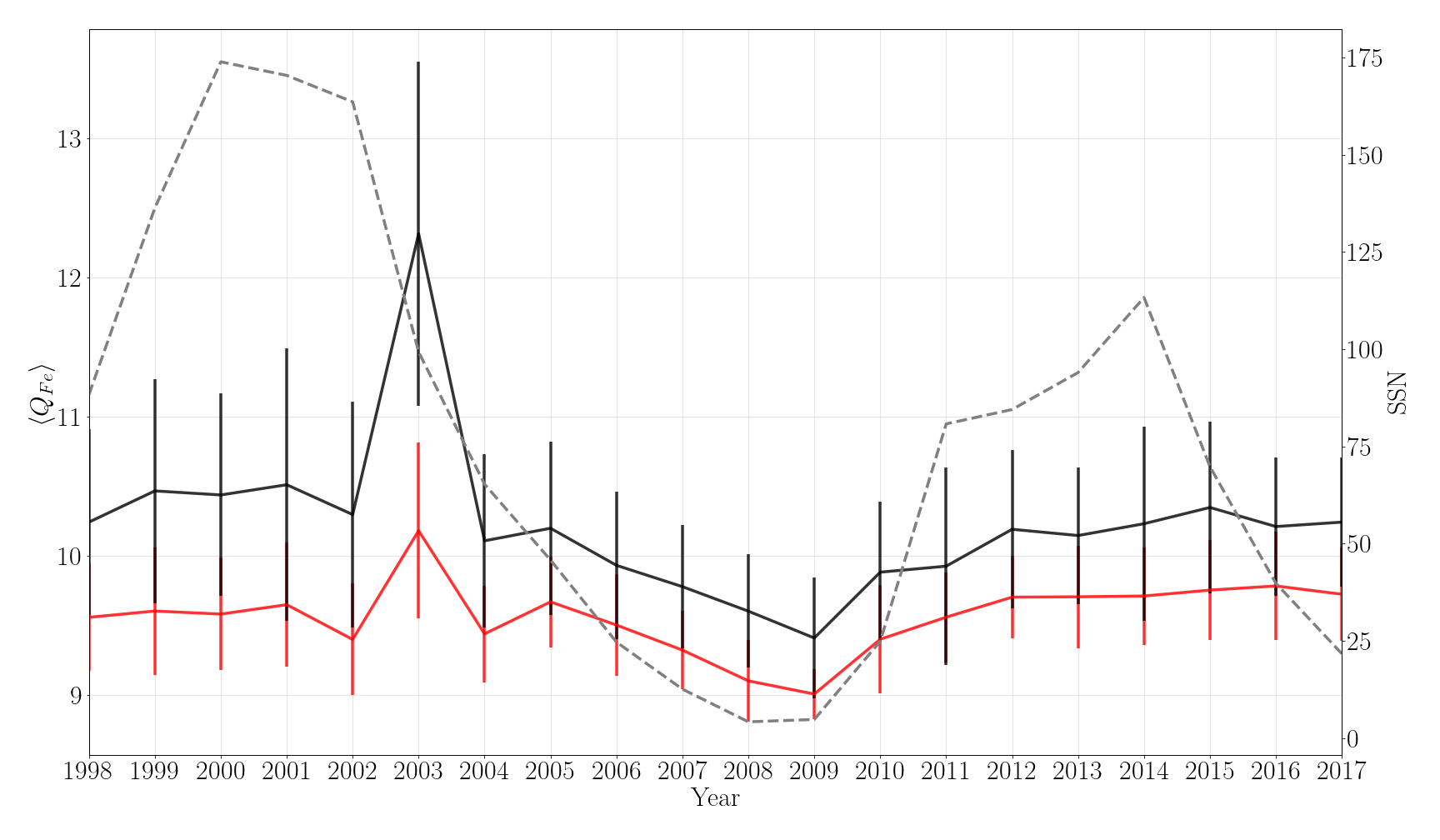}
    \caption{Position of the peaks from the bi-Gaussian fitting for each year. Red (black) line represents the first (second) Gaussian. The RMS width of the corresponding Gaussian ($w$) has been considered as the uncertainty in the $p$ parameter. The grey dashed line corresponds to the yearly SSN.}    \label{fig:position_evolution_ssn}
\end{figure}

\begin{figure}[h!]
	\includegraphics[width=0.96\columnwidth]{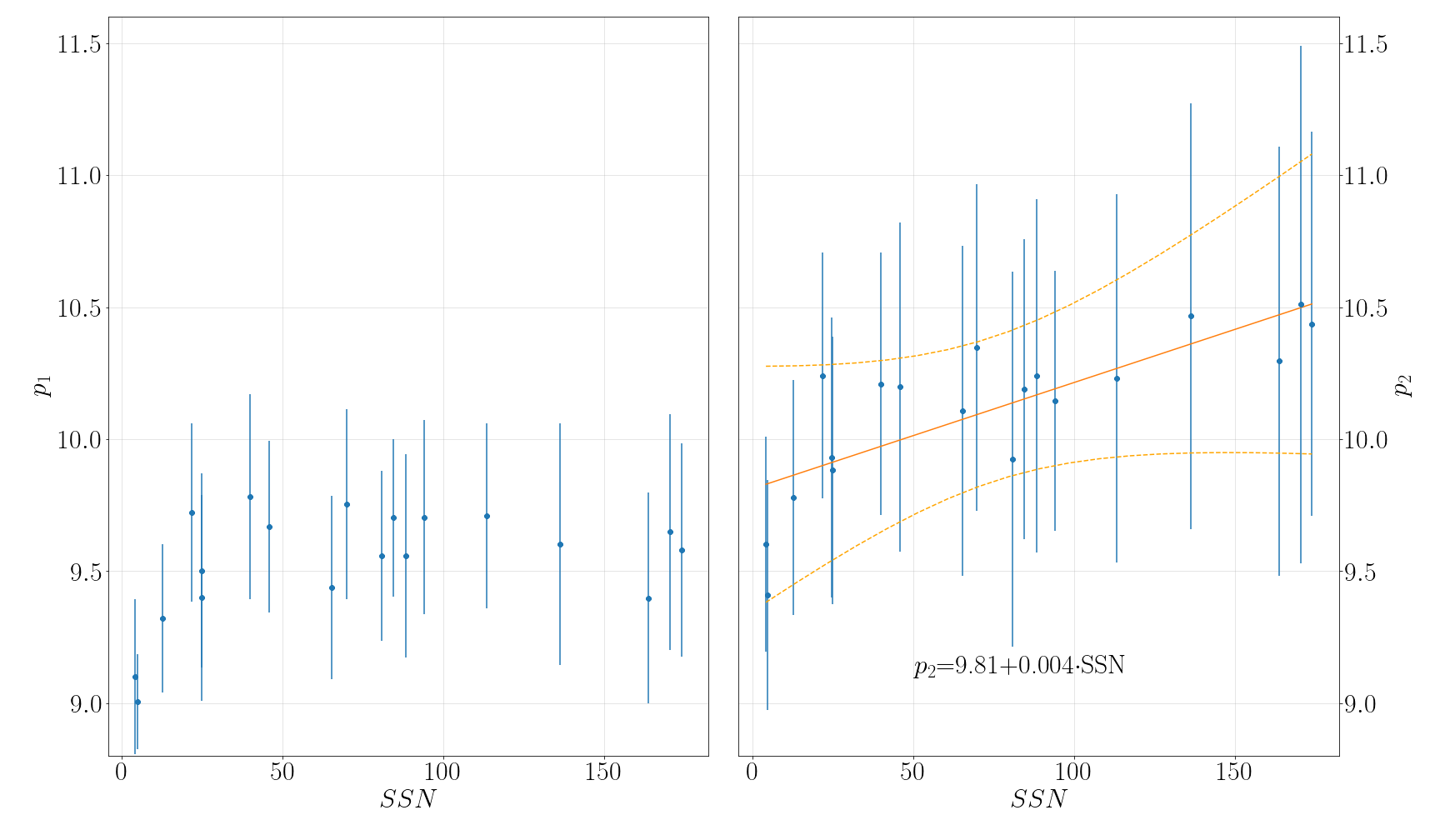}
    \caption{Scatter plots of $p_{1}$ (left) and $p_{2}$ (right) vs SSN. In the right panel, the orange solid line corresponds to the linear regression and the orange dashed lines are the 99\% confidence intervals.}
    \label{fig:position_evolution_ssn_no_2003}
\end{figure}

Considering this result, we study the relationship between the bimodal $\langle Q_{Fe}\rangle$ distribution function and the solar cycle, using the sunspot number (SSN) as a proxy for solar activity. 
Figure \ref{fig:position_evolution_ssn} compares the position of the peaks from the bi-Gaussian fitting and the SSN over time. Year 2003 diverges from the trend of the rest of the period analyzed because of a very high activity with a major contribution from the fast wind. Indeed, \cite{Larrodera_2020_Bimodal_solar_wind} described it as an unsusual year dominated by fast solar wind and a large number of CMEs. Therefore, we have avoided including year 2003 to search for any relationship with the solar cycle. 
Figure \ref{fig:position_evolution_ssn_no_2003} shows the scatter plots of positions $p_{1}$ and $p_{2}$ versus SSN (excluding year 2003).
We cannot perceive any clear correlation between $p_{1}$ and the SSN, but $p_{2}$ clearly increases as SSN increases. Indeed, a Pearson correlation coefficient $r\left(p_{2},SSN\right)=0.77$ (excluding year 2003) indicates a strong linear correlation between $p_{2}$ and the SSN. 
On the contrary, $r\left(p_{1},SSN\right)=0.40$, confirms the weak correlation observed between $p_{1}$ and the SSN. 

\vspace{-1.15cm}

\newpage
\subsection{Removing Outliers. Analyzing the Bulk Solar Wind}

The previous section analyzes the distribution function of the solar wind as a whole data set with all data contributing to the bulk solar wind. However, we appreciate in Figure \ref{fig:monthly_gauss} some months with extremely high values of $\langle Q_{Fe}\rangle$, which may be associated with transient events. Indeed, \cite{Lepri_2001_ICME_QFe} and \cite{Lepri_2004_ICME_Composition} considered large $\langle Q_{Fe}\rangle$ as a sufficient signature of ICMEs. Moreover, Figure 1 in \cite{Lepri_2004_ICME_Composition} shows that the probability that one would find a certain average charge state in the normal solar wind is zero from $\langle Q_{Fe}\rangle=12$ on. Checking the results from the previous bi-Gaussian fitting to the whole data set (Figure \ref{fig:qfe_all}), we notice that $\langle Q_{Fe}\rangle=12$ corresponds to $p_{2}+3\sigma_{2}$, and therefore values of $\langle Q_{Fe}\rangle > 12$ can be considered as outliers of the bulk solar wind.
As our first goal is to study the bulk solar wind, we proceed to remove all these outliers from the whole data sample. 

\begin{figure}[h]
	\includegraphics[width=0.96\columnwidth]{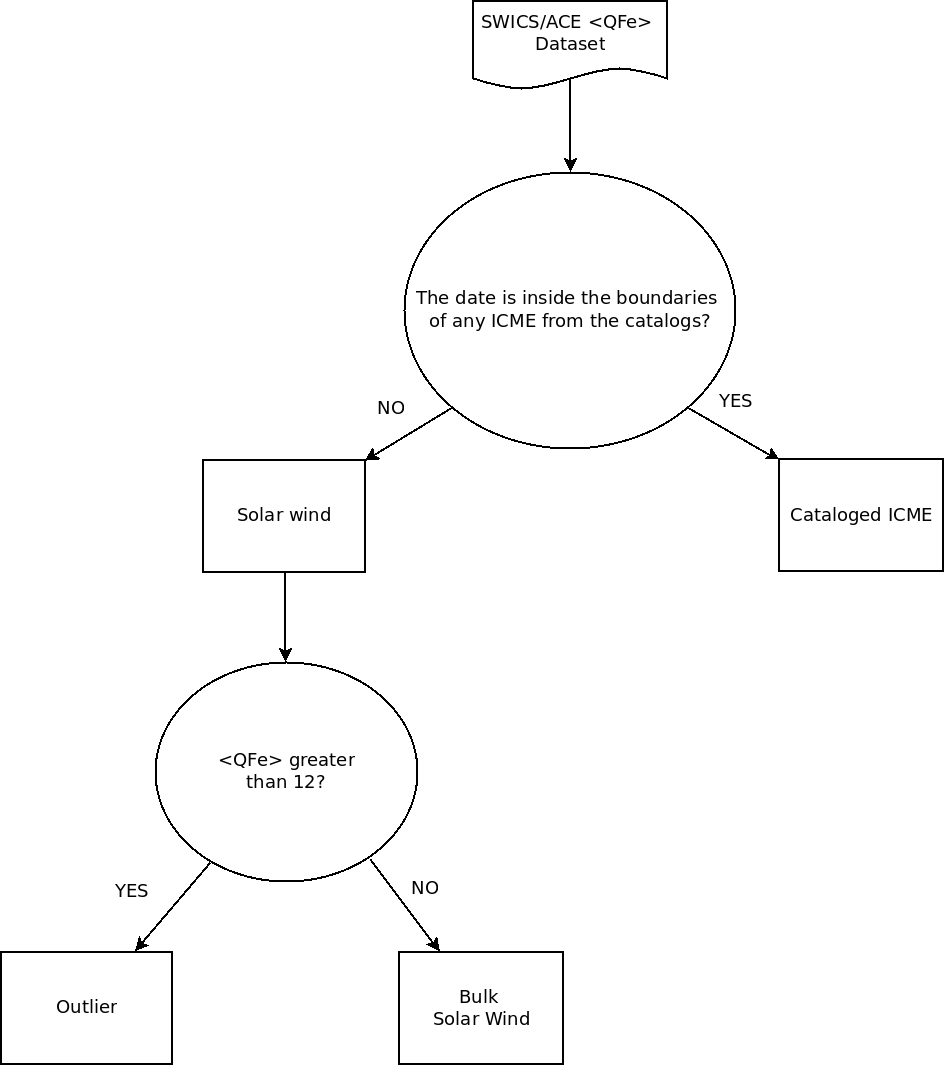}
    \caption{Workflow showing the process to separate the bulk solar wind data from the outliers and the ICMEs previously identified.}
    \label{fig:workflow}
\end{figure}

We also remove from the whole data sample the ICMEs previously identified. For this purpose we considered the events listed in the following ICME catalogs: Richarson and Cane \urlstyle{same}\url{(http://www.srl.caltech.edu/ACE/ASC/DATA/level3/icmetable2.htm)}, \\ \cite{Jian_2006_ICME_Catalog}, \cite{Jian_2011_ICME_Catalog} and NASA Wind catalog \urlstyle{same}\url{(https://wind.nasa.gov/cycle22.php)}. Figure \ref{fig:workflow} shows the workflow of the process. 

After this process, the reduced data sample, representing the bulk solar wind, is ready for analysis. Thus, we go ahead fitting the bi-Gaussian function to the PDF of the reduced data set of $\langle Q_{Fe}\rangle$, which represent 89 \% of the whole.

\begin{figure}[h]
	\includegraphics[width=0.96\columnwidth]{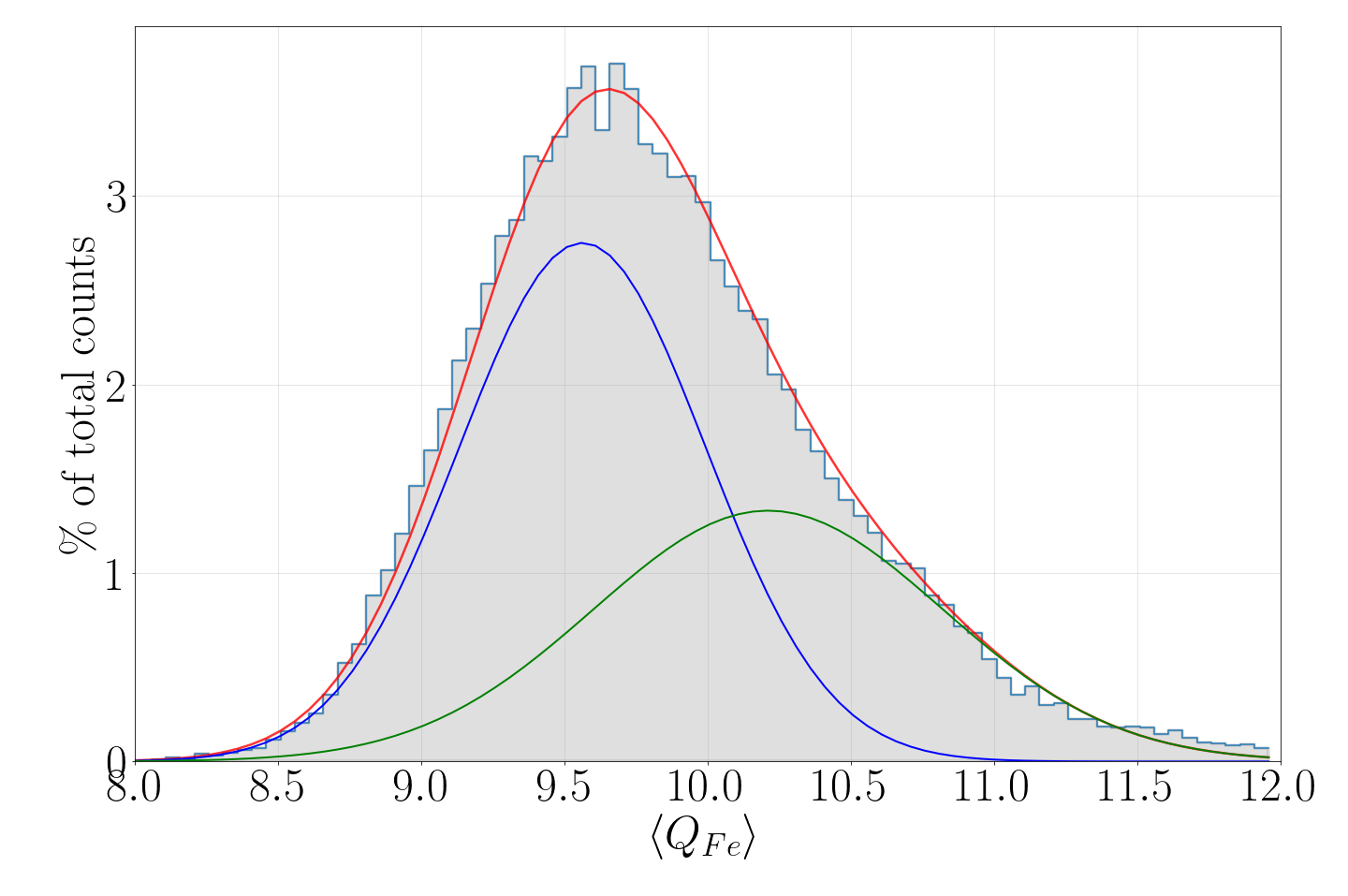}
   \caption{Empirical distribution function of the $\langle Q_{Fe}\rangle$ for the reduced sample (shadowed as grey) and fitting to a bi-Gaussian function (red). Green and blue lines correspond to the single Gaussian curves.}
    \label{fig:qfe_all_no_cme}
\end{figure}

Figure \ref{fig:qfe_all_no_cme} shows the empirical distribution function of the $\langle Q_{Fe}\rangle$ for the reduced data sample and the fitting to the bi-Gaussian function. A Pearson-$\chi^{2}$ value of $0.01$ shows a great fit for the bulk solar wind sample.
Now, the values of $\langle Q_{Fe}\rangle$ are spread around $p_{1}=9.6$ and $p_{2}=10.2$, almost around the same values as before, but with smaller uncertainties. 

\begin{figure}[h]
	\includegraphics[width=0.96\columnwidth]{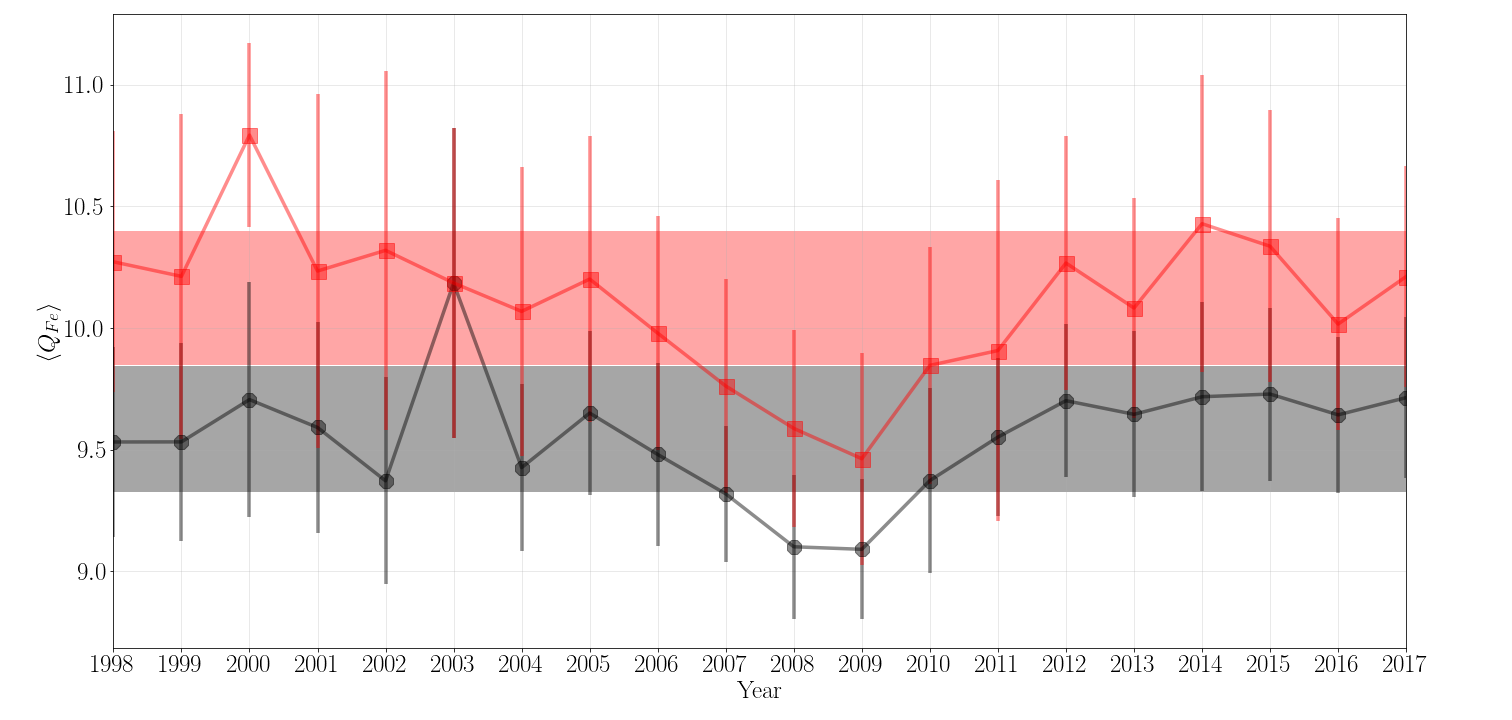}
    \caption{Position of the peak of the first (black) and second (red) Gaussian distribution function from the yearly fitting. The shadowed areas are the weighted average position $\pm$ $\sigma_{w}$.}
    \label{fig:qfe_positions}
\end{figure}

Figure \ref{fig:qfe_positions} represents the position of the peaks from the bi-Gaussian fitting for each year of the bulk solar wind data set. 
The red (black) shadowed area corresponds to the weighted average position of the peak $p_{1}$ ($p_{2}$) $\pm \sigma_{w}$.
Although both colored regions in the plot are separated, reinforcing a bimodal distribution function for $\langle Q_{Fe}\rangle$ at 1 AU, the position of one of the peaks appears in the region corresponding to the other peak for some years. This happens, not only for year 2003, which we labeled as anomalous above, but also for years 2007, 2008 and 2009.  

\begin{figure}[h]
	\includegraphics[width=0.96\columnwidth]{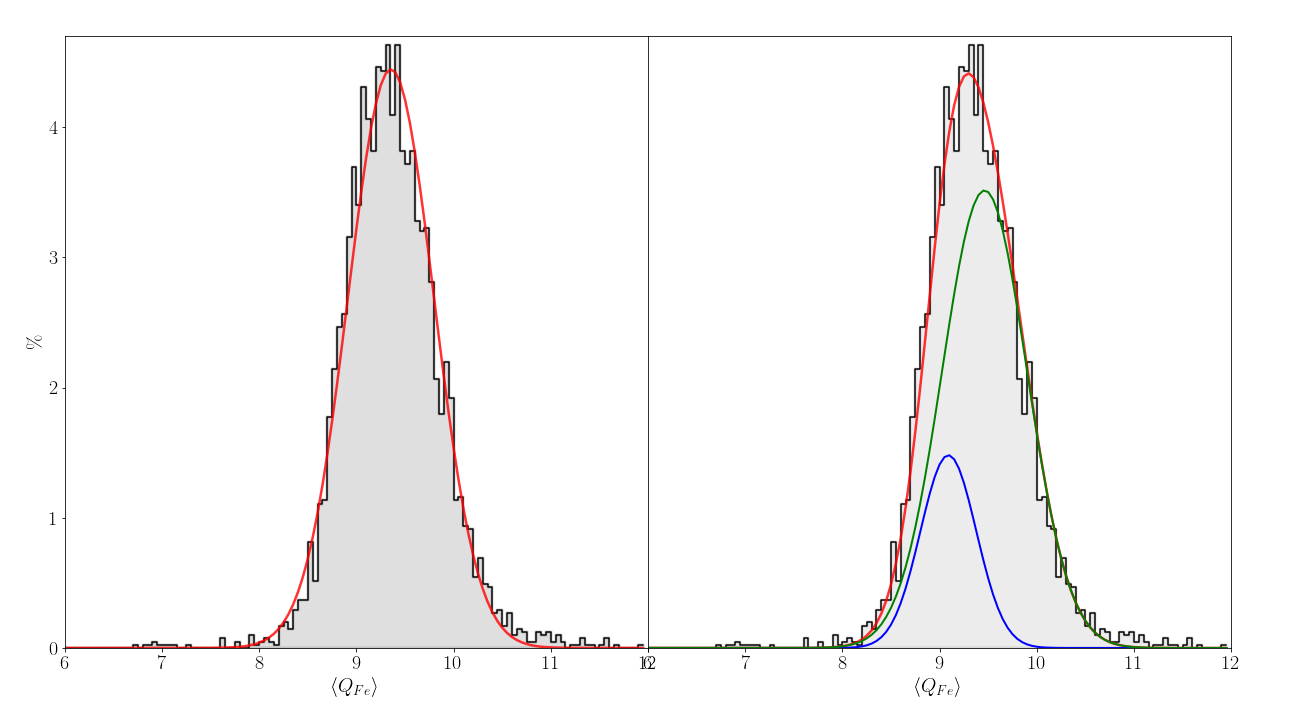}
   \caption{Empirical distribution function of the $\langle Q_{Fe}\rangle$ for the reduced sample of year 2009 with a single Gaussian fitting (left) and bi-Gaussian fitting (right) with the corresponding single curves in blue and green.}
    \label{fig:fitting_1_2G}
\end{figure}

\begin{figure}[h]
	\includegraphics[width=0.96\columnwidth]{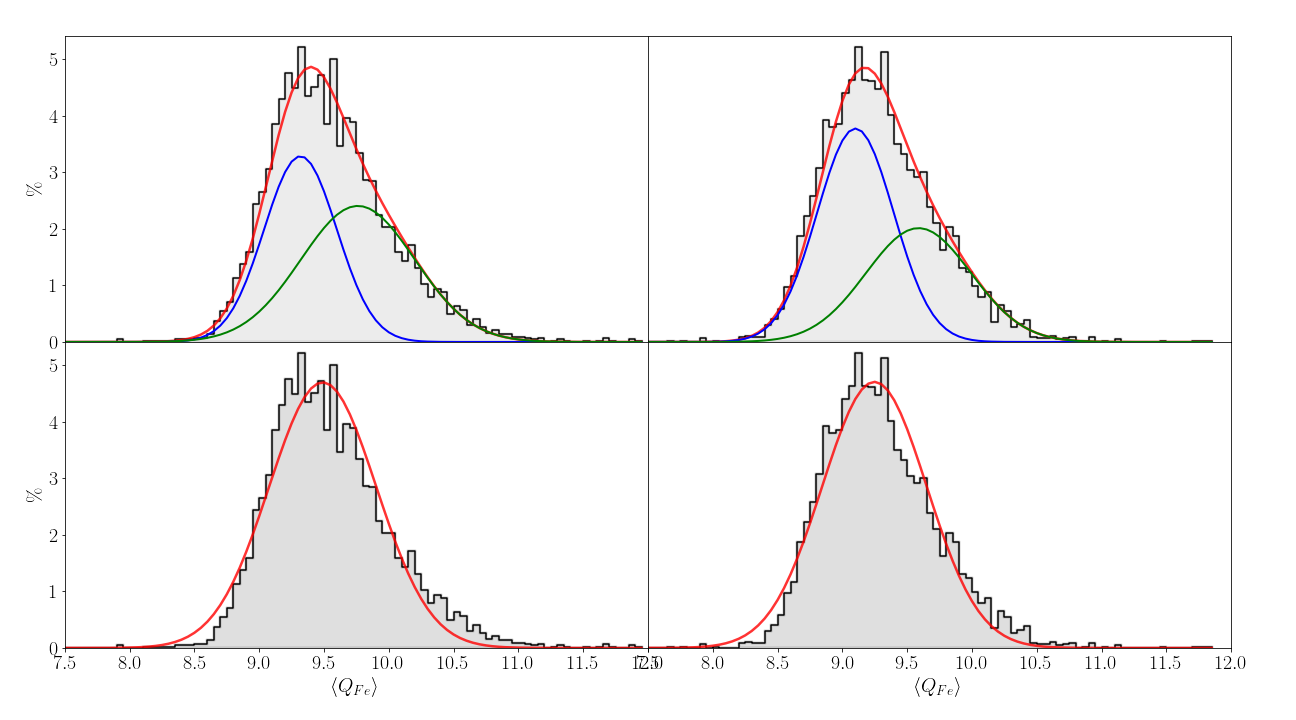}
    \caption{Empirical distribution function of the $\langle Q_{Fe}\rangle$ for the reduced sample (shadowed in grey). Left (right) panels correspond to year 2007 (2008). Top panels show the bi-Gaussian fittings (in red) with the corresponding single curves in blue and green. The red curve in bottom panels corresponds to the single Gaussian fittings.}
    \label{fig:fitting_1_2G_2007_2008}
\end{figure}

\cite{Larrodera_2020_Bimodal_solar_wind} explained that in 2009 the slow solar wind is highly dominant over the fast wind.
In this situation, with predominance of slow solar wind, the contribution of the fast solar wind can be discarded and the bi-Gaussian fitting can be replaced by a single Gaussian fitting.
This is shown in Figure \ref{fig:fitting_1_2G}, where we compare the bi-Gaussian and single Gaussian fitting for 2009. The single Gaussian fitting provides a good approximation for the data set, and adding a second one (i.e. using bi-Gaussian fitting) will not provide a better fitting. Indeed, the first Gaussian has a  smaller height when compared to the second Gaussian, therefore its contribution can be discarded as this has no physical meaning.
Nevertheless, this is not the case for years 2007 and 2008. Figure \ref{fig:fitting_1_2G_2007_2008} shows a clear deviation from the single Gaussian fitting of the PDF of years 2007 and 2008 due to the heavy tail. This problem is solved by the bi-Gaussian function. 
Indeed, the displacement of some years from the corresponding shadowed areas in Figure \ref{fig:qfe_positions} may be related to the solar cycle.

\begin{figure}[ht!]
	\includegraphics[width=0.96\columnwidth]{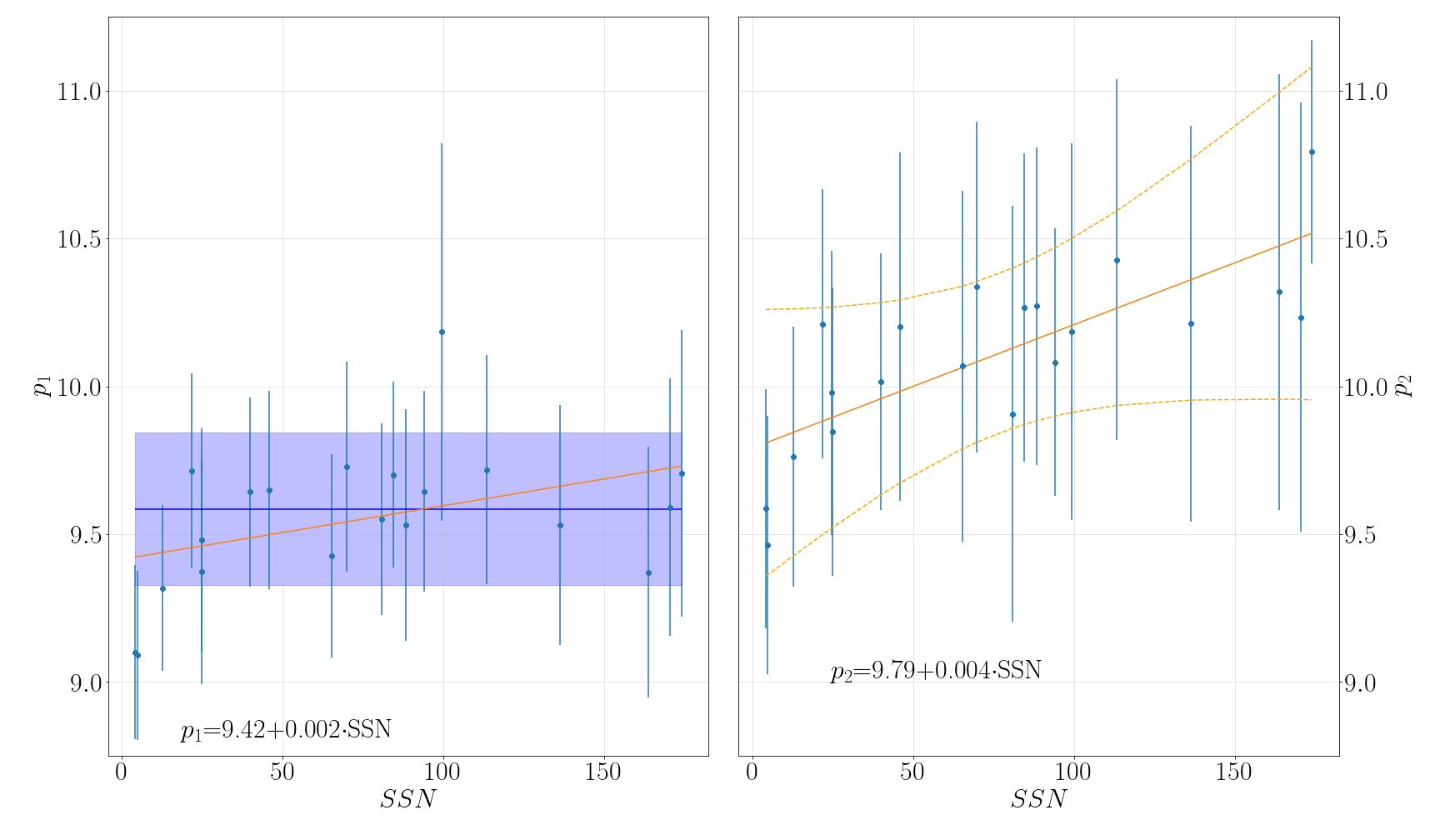}
   \caption{Scatter plots of $p_{1}$ (left) and $p_{2}$ (right) vs. SSN. The blue shadowed area represents the weighted average of $p_{1} \pm\sigma_{w}$.
    The orange solid lines correspond to the linear regression and the orange dashed lines in the right panel are the 99\% confidence intervals.}
    \label{fig:position_evolution_ssn_no_icme}
\end{figure}

Figure \ref{fig:position_evolution_ssn_no_icme} shows the scatter plot of $p_{1}$ and $p_{2}$ versus the SSN for the reduced sample. The values of the Pearson correlation coefficient $r\left(p_{1},SSN\right)=0.41$ and $r\left(p_{2},SSN\right)=0.76$ are similar to those obtained before, showing a strong (weak) linear relationship of $p_{2}$ ($p_{1}$) with the solar cycle, but this time including year 2003 in the sample. 
Indeed, the linear relationship of $p_{1}$ with the solar cycle is so weak that the interval centered in the weighted average with an uncertainty of $\pm\sigma_{w}$ ($p_{1}=9.6\pm0.3$) includes the regression line.
Note also that the $y$-intercept of the linear regression between $p_{2}$ and the SSN ($p_{2}=9.8$) is included in the interval 
$p_{1}=9.6\pm0.3$. This result allows us to recognize that the type of wind labeled with the subindex 1 corresponds to the slow wind. Indeed, in the case of no sunspot (i.e. at the $y$-intercept) no relevant coronal holes out of the poles are foreseen (as expected in the solar minimum) and therefore the only type of wind will be the slow one. This case would be similar to year 2009, where a single Gaussian function is enough to describe the whole bulk solar wind. Thus, for SSN=0, both single Gaussian curves are expected to coincide.

\newpage
\section{Are ICMEs the Outliers?} \label{sec:outliers_ICMEs}
The whole sample of $\langle Q_{Fe}\rangle$ data from SWICS/ACE has been split into three sets after the workflow in Figure \ref{fig:workflow}: bulk solar wind (or reduced sample), ICMEs previously identified, and  outliers. 
The set of outliers includes all data where $\langle Q_{Fe}\rangle > 12$, which where not previously identified as ICME material. We ask now, which type of solar wind are the outliers? Where do they come from? Considering the physical processes happening in the solar atmosphere, they cannot be part of the slow solar wind nor of the fast wind coming from coronal holes. Indeed, \cite{Lepri_2001_ICME_QFe} show that the $\langle Q_{Fe}\rangle$ of the bulk solar wind are typically around 9 to 11. 

\begin{figure}[h]
	\includegraphics[width=0.96\columnwidth]{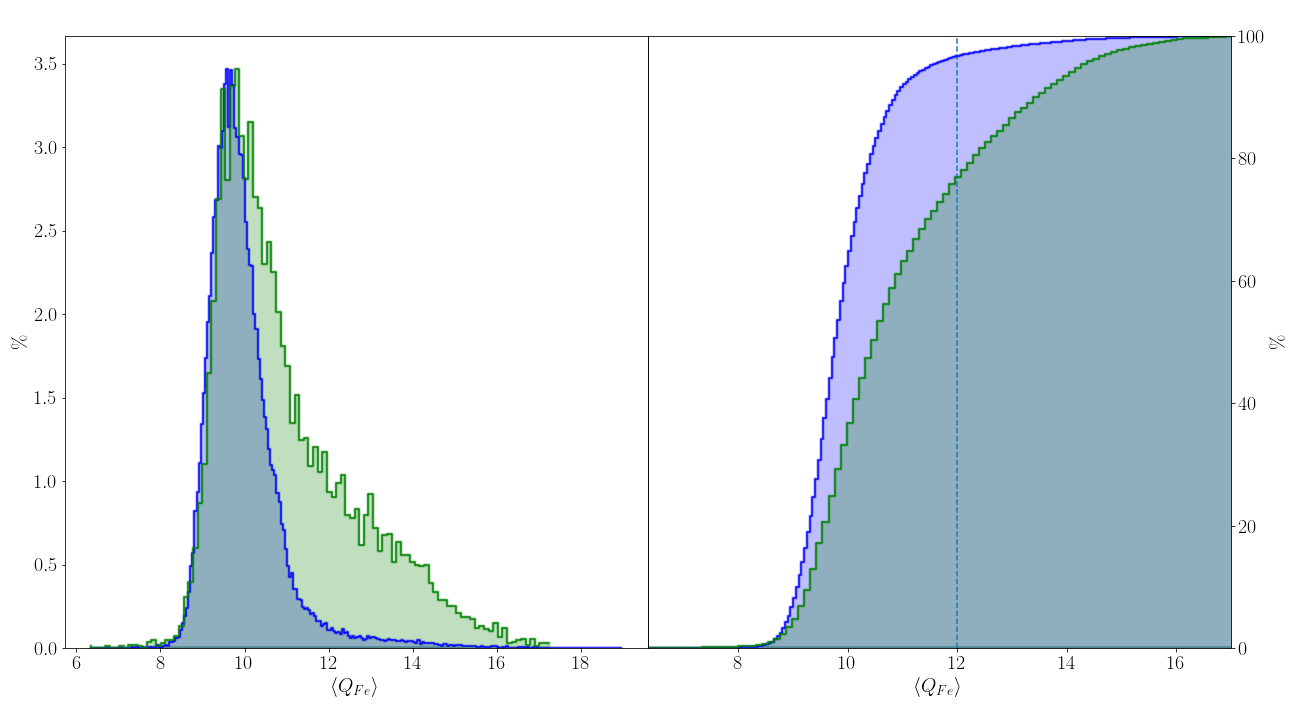}
    \caption{(Left figure) Distribution function of $\langle Q_{Fe}\rangle$ for the whole ACE data set (blue) and for the ICMEs from the catalogs (green). The distribution function of the ICMEs has been rescaled for comparison. 
    (Right figure) Cumulative distribution function of $\langle Q_{Fe}\rangle$ for the whole ACE data set (blue) and for the ICMEs from the catalogs (green).}
    \label{fig:pdf_function}
\end{figure}

The right panel in Figure \ref{fig:pdf_function} compares the cumulative distribution function of $\langle Q_{Fe}\rangle$ for the whole ACE data set (in blue) and for the ICMEs from the catalogs described above (in green). It can be appreciated that more than 95\% (exactly 96.7\%) of the values of the $\langle Q_{Fe}\rangle$ are below 12 for the whole solar wind sample, reinforcing the result by \cite{Lepri_2004_ICME_Composition}. This value drops to 76.9\% for the data sample from the ICME catalogs and, although just 23.1\% of the values of $\langle Q_{Fe}\rangle$ from the ICMEs are larger than 12, the greater-than-12 part of the distribution function is very distinct to that of the whole sample (left panel Figure \ref{fig:pdf_function}). The less-than-12 part of the distribution function of the ICME values may be related to ICMEs with no large flaring activity (avoiding large ionization states of Fe) and/or to ICMEs containing also normal solar wind. Indeed, \cite{Richardson_2005} noted that different signatures of an ICME may not occur exactly concurrently. Being aware of this problem,  \cite{Jian_2006_ICME_Catalog} included in their ICME intervals the shock (if it occurs), sheath pile-up region, and the ejecta.

Certainly, high charge states arise because high temperatures in the solar corona, associated with the initiation of CMEs, ionize the material ejected into the solar wind. Thus, large deviations from typical values of $\langle Q_{Fe}\rangle$ are expected to be associated with ICMEs. Therefore, the goal of this section is to analyze the solar wind parameters during outliers with the aim of identifying any signature of ICME material, other than enhanced $\langle Q_{Fe}\rangle$. 
For this purpose, we have arranged the outliers as a list of events, considering an event when $\langle Q_{Fe}\rangle>12$ during at least 10 hours consecutively. The result includes 27 events, which are listed in Table \ref{tab:top_events}. The times listed in Table \ref{tab:top_events} are accurate to within 2 hours (the time resolution of the data set). 

\begin{table}
\caption{List of outlier events}
\label{tab:top_events}
\begin{tabular}{c c c c} 
\hline \hline
$\langle Q_{Fe}\rangle > 12$ Start & $\langle Q_{Fe}\rangle > 12$ End & Type of\\ 
year month day UT  & year month day UT & event \\ \hline
1999 01 09 18h &1999 01 10 04h &New \\
1999 11 15 00h &1999 11 15 16h &Extended\\
2000 03 11 05h &2000 03 11 13h &New\\
2000 07 17 09h &2000 07 17 23h &Extended\\
2000 12 24 06h &2000 12 24 14h &Extended\\
2002 06 27 12h &2002 06 27 22h &New\\
2002 09 23 06h &2002 09 23 16h &Extended\\
2002 09 26 06h &2002 09 26 14h &New\\
2003 04 28 19h &2003 04 29 05h &New\\
2003 07 08 17h &2003 07 09 03h &Extended\\
2003 07 09 11h &2003 07 09 19h &Extended\\
2003 07 10 23h &2003 07 11 09h &Extended\\
2003 10 30 11h &2003 10 31 03h &Extended\\
2003 11 02 00h &2003 11 02 08h &Extended\\
2005 01 20 03h &2005 01 20 13h &Extended\\
2005 07 12 15h &2005 07 13 01h &Extended\\
2005 12 07 10h &2005 12 09 08h &New\\
2010 02 15 01h &2015 02 15 11h &New\\
2011 02 18 05h &2011 02 18 21h &Extended\\
2011 08 06 08h &2011 08 06 20h &Extended\\
2012 06 18 02h &2012 06 18 10h &Extended\\
2013 05 22 06h &2013 05 22 14h &New\\
2013 10 27 19h &2013 10 28 03h &New\\
2015 01 29 14h &2015 01 30 06h &New\\
2015 02 04 20h &2015 02 05 12h &New\\
2015 03 05 04h &2015 03 05 16h &New\\
2016 07 22 15h &2016 07 24 03h &Extended\\ \hline
\end{tabular}
\end{table}

The events have been classified into two groups (see column 3 in Table \ref{tab:top_events}): 'Extended' and 'New'. Those cataloged as 'Extended' are events where an extension of the boundaries of an already identified ICME will include the outlier event. When there is no identified ICME close to the outlier event, it is labeled as 'New'.

\begin{figure}[h!]
{\includegraphics[width=\columnwidth]{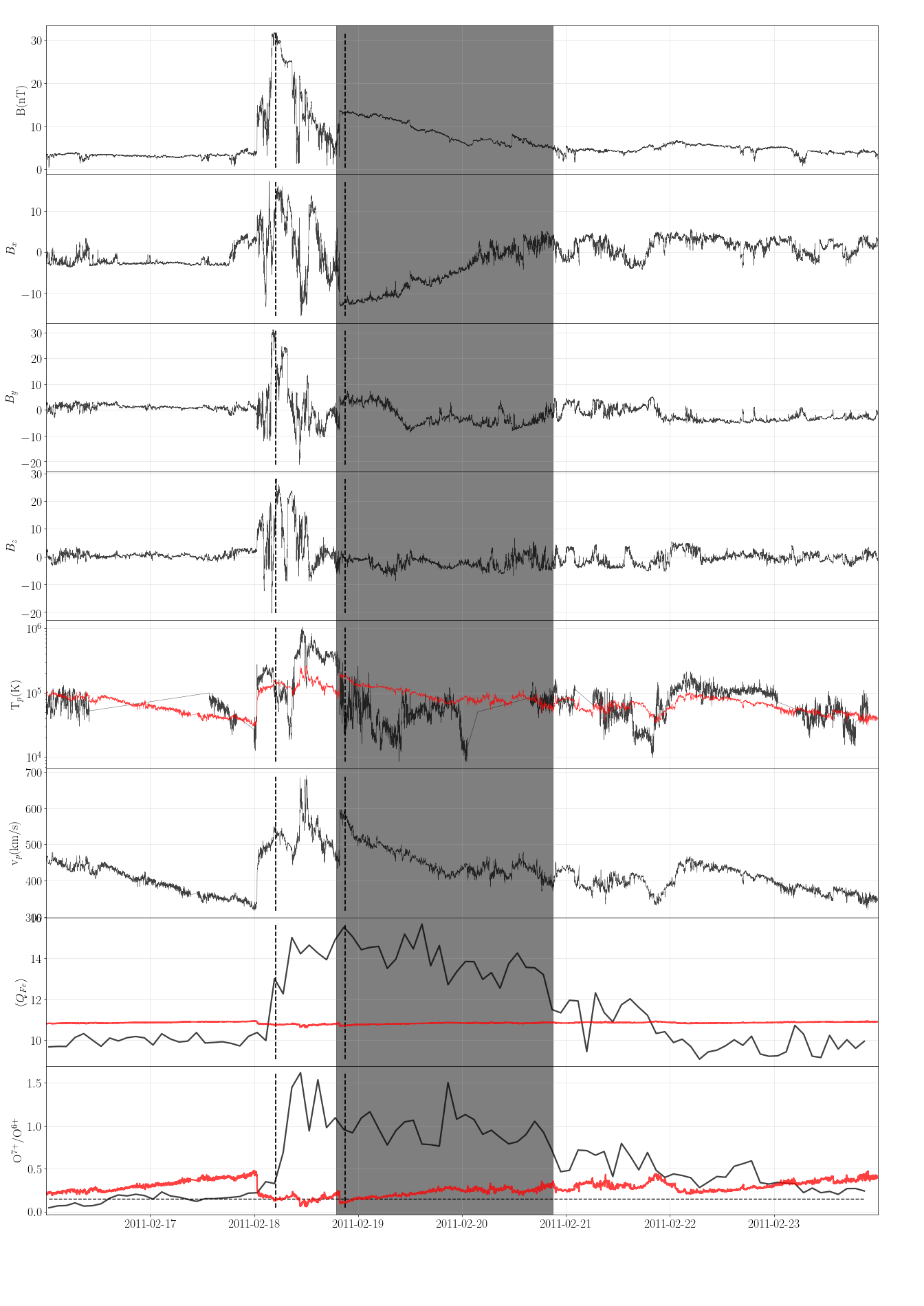}}
\caption{Solar wind parameters from 16 to 23 February 2011. From top to bottom: magnetic field strength ($B$), magnetic field components ($B_{x}$, $B_{y}$, $B_{z}$), proton temperature ($T_{p}$), proton speed ($v_{p}$), average iron charge state $\langle Q_{Fe} \rangle$ and oxigen ratio $O^{7+}/O^{6+}$}
\label{fig:event_19}
\end{figure}

\begin{figure}[h!]
{\includegraphics[width=\columnwidth]{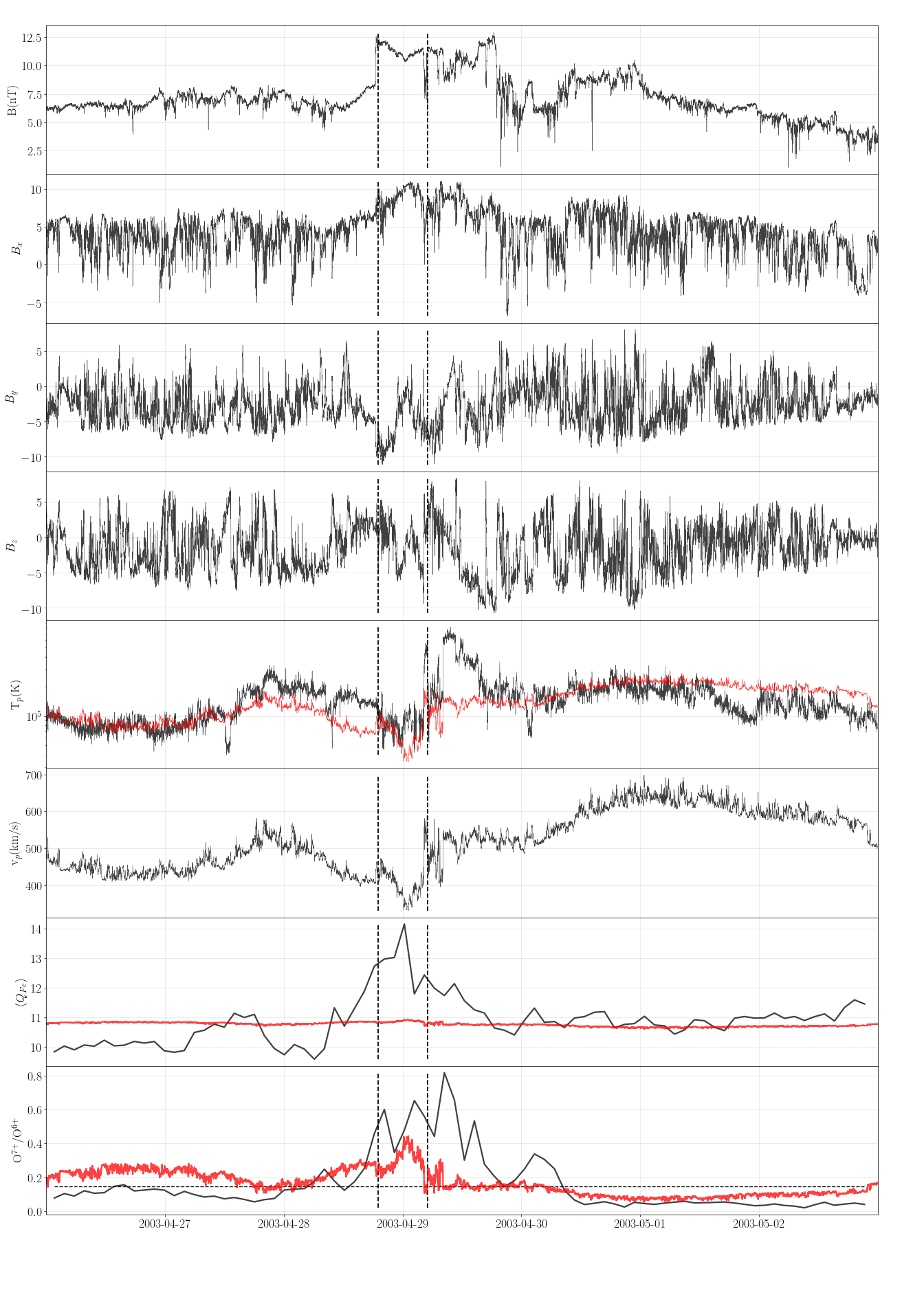}}
\caption{Solar wind parameters from 26 April to 2 May 2003. From top to bottom: magnetic field strength ($B$), magnetic field components ($B_{x}$, $B_{y}$, $B_{z}$), proton temperature ($T_{p}$), proton speed ($v_{p}$), average iron charge state $\langle Q_{Fe} \rangle$ and oxigen ratio $O^{7+}/O^{6+}$}
\label{fig:event_9}
\end{figure}

Figure \ref{fig:event_19} shows the solar wind parameters measured by ACE during the event on 18 February 2011, as an example of an 'Extended' event. From top to bottom Figure \ref{fig:event_19}  shows the magnetic field strength ($B$) and its GSM components ($B_{x}$, $B_{y}$, $B_{z}$) from the Magnetic Field Experiment instrument (MAG/ACE); the proton temperature ($T_{p}$), and the proton speed ($v_{p}$) from the Solar Wind Electron, Proton, and Alpha Monitor instrument (SWEPAM/ACE), and the $\langle Q_{Fe} \rangle$, and $O^{7+}/O^{6+}$ ratio from SWICS/ACE, from 16 to 23 February 2011. Superimposed on the observed values of $T_{p}$, $\langle Q_{Fe} \rangle$, and $O^{7+}/O^{6+}$ ratio, as red solid lines, are the expected values according to \cite{Richardson_1995_CME_Temperature} and \cite{Richardson_2004_ICME_Identification_Composition}.
The horizontal black dashed line in the bottom panel indicates the reference $O^{7+}/O^{6+}= 0.145$. \cite{Zhao_2009_composition} and  \cite{Heidrich_2016_Composition} state that values below this threshold correspond to fast streams coming from coronal holes.




The shadowed area in Figure \ref{fig:event_19} corresponds to the ICME interval as identified in the catalogs previously detailed. As there is not an agreement on the boundaries of this event (as in many others), we have drawn the shadowed area so that it covers the greater interval considering the boundaries of different catalogs. 
In Figure \ref{fig:event_19}, the start date of the shadowed area comes from the Richardson and Cane catalog and the end date from the Wind catalog. The dashed vertical lines indicate the boundaries of the region where $\langle Q_{Fe} \rangle > 12$. Values of $\langle Q_{Fe} \rangle$ over 14 reached in the interval of the outlier are difficult to explain if this solar wind does not comes from a CME.
The larger values of the $O^{7+}/O^{6+}$ ratio relative to the expected one also support the hypothesis of CME material in this interval.
Between the front boundary of the outlier (first dashed line) and the beginning of the shadowed area, the magnetic field strength reached more than 30 nT, about more than twice the one in the shadowed area. This high value in the magnetic field strength may be related to the interaction between a halo CME launched on 15 February 2011 at 02:24UT and a previous slower halo CME first appearing on the second telescope of the Large Angle and Spectrometric Coronagraph (C2/LASCO) on 14 February 2011 at 18:24UT. In this scenario, the outlier interval corresponds to the first ICME which is compressed by the second one, appearing in the shadowed area. As in the event analyzed by \cite{Cid_2016_ICME_Boundaries}, during this event the boundary identification from magnetic signatures does not agree with those from other signatures. This discrepancy can be explained as due to the interaction of transients, which modifies the magnetic topology of ICMEs. 

Detailed analysis of the events in Table \ref{tab:top_events} guide us to conclude that 14 additional events present similar features to those described above during the event on 18 February 2011, where an extension of the ICME boundaries allow to consider the outliers as ICME material. But what about the other 12 events (almost half of the sample)? Values of $\langle Q_{Fe} \rangle$ over 12 for more than 10 hours are difficult to explain if the solar wind transient is not an ICME. In some cases the ICME might have been missed in the catalogs due to data gap in any relevant magnitude used during the identification process. For example, a data gap in temperature of about one day appears during the event on 11 March 2000, but there are several events without data gaps. The event on 28\,--\,29 April 2003 is an example of the 'New' events without data gap. Solar wind parameters during that event appear in Figure \ref{fig:event_9} with the same format as in Figure \ref{fig:event_19}. The interval of the outlier coincides with a region where the magnetic field is enhanced up to more than 10 nT and the proton temperature is below the expected one, according to \cite{Richardson_1995_CME_Temperature}. The outlier interval is surrounded by solar wind with high speed (500\,--\,600 km/s) and with both magnetic field vector and velocity highly fluctuating, as expected from the fast wind coming from a coronal hole. Indeed, the $O^{7+}/O^{6+}$ ratio shows values below 0.145, reinforcing the signatures of fast streams. In this scenario, the large discontinuities in the magnetic field strength at the boundaries of the outlier appear as signatures of the interaction of the surrounding fast wind with an ICME (identified as a 'New' outlier event). Moreover, before and after the outlier, the value of $\langle Q_{Fe}\rangle$ is around 10 and 11, respectively, which is compatible with the type of wind labeled with subindex 2 (see Figure \ref{fig:qfe_all_no_cme}.)

\section{Summary and Conclusions} 
\label{sec:conclusions}

Here we show that the bi-Gaussian function reproduces the  empirical distribution function of the $\langle Q_{Fe} \rangle$ of the bulk solar wind at 1 AU. This bimodal wind presents two components which spread around $\langle Q_{Fe} \rangle = 9.6$ and 10.3, with the last one strongly dependent on the solar cycle. These two components are supposed to be associated with slow and fast wind. Nevertheless, 
Figure \ref{fig:position_evolution_ssn_no_2003} and Figure \ref{fig:position_evolution_ssn_no_icme} demonstrate that $p_{2}$ is related with the solar cycle. The relationship of the number of active regions and the SSN is undeniable. Thus, this result supports the results from \cite{Neugebauer_2002_Solar_Wind} identifying two solar sources of the fast solar wind: the coronal holes and the active regions.

Our results are obtained from the analysis of the whole data set of $\langle Q_{Fe}\rangle$ from SWICS/ACE from 1998 to 2017, after separating the bulk solar wind from the transients. The transients include, not only the events previously identified as ICMEs in different sources, but also those intervals where  $\langle Q_{Fe} \rangle > 12$, which we labeled as outliers. 
The analysis of the whole data sample shows that the threshold set by \cite{Lepri_2004_ICME_Composition} in $\langle Q_{Fe}\rangle$=12 for the ICMEs, corresponds to  a deviation of 3$\sigma$ from the position of the peak of the second Gaussian distribution, supporting the establishment of this threshold as robust for ICMEs, as ICMEs are the unique outlier of the bulk solar wind known to date.

We also provide a catalog of 27 outliers where the condition $\langle Q_{Fe} \rangle > 12$ is maintained at least for 10 hours. From the analysis of the different solar wind parameters around the events of the catalogue, we report that half of the events can be considered as part of an already identified ICME or a group of ICMEs. We identify new ICME events in the other half of the catalog. These events may be missing in the existing catalogs because of a data gap in one of the solar wind parameters commonly used to identify ICMEs or due to an anomalous behavior of some of the parameters due to interaction of the CME material with the surrounding solar wind or even with other ICMEs. From our results we strongly support that $\langle Q_{Fe} \rangle > 12$ is a sufficient signature to identify ICMEs in the solar wind and the most convenient signature to identify its boundaries. Future work will be dedicated to understand whether the lower-than-12 values of  $\langle Q_{Fe} \rangle$ in ICMEs are associated with their solar origin (i.e. non large flaring) or to the boundary identifications.

\section{Acknowledgements}
  This work was supported by the MINECO project AYA2016-80881-P (including FEDER funds). We thank the SWICS, SWEPAM and MAG instruments teams and the ACE Science Center for providing the ACE data. We acknowledge WDC-SILSO, Royal Observatory of Belgium, Brussels for providing the sunspot number. We also acknowledge the information from the CME catalog generated and maintained at the CDAW Data Center by NASA and The Catholic University of America in cooperation with the Naval Research Laboratory. SOHO is a project of international cooperation between ESA and NASA. The authors want to thank an anonymous reviewer for the useful comments.
Disclosure of Potential Conflicts of Interest: The authors declare that there are no conflicts of interest.

\bibliographystyle{apalike}  
\bibliography{references}

\begin{thebibliography}{}

\bibitem[Bale et~al., 2019]{Bale_2019_Solar_wind_CH}
Bale, S.~D., Badman, S.~T., Bonnell, J.~W., Bowen, T.~A., Burgess, D., Case,
  A.~W., Cattell, C.~A., Chandran, B. D.~G., Chaston, C.~C., Chen, C. H.~K.,
  et~al. (2019).
\newblock Highly structured slow solar wind emerging from an equatorial coronal
  hole.
\newblock {\em Nature}, pages 1--6.

\bibitem[Banaszkiewicz et~al., 1997]{Banaszkiewicz_1997_Fast_SW}
Banaszkiewicz, M., Czechowski, A., Axford, W.~I., McKenzie, J.~F., and
  Sukhorukova, G.~V. (1997).
\newblock {The Fast Solar Wind and its Source Region}.
\newblock In {Wilson}, A., editor, {\em Correlated Phenomena at the Sun, in the
  Heliosphere and in Geospace}, volume 415 of {\em ESA Special Publication},
  page~17.

\bibitem[Burlaga and King, 1979]{Burlaga_1979_IMF}
Burlaga, L.~F. and King, J.~H. (1979).
\newblock {Intense interplanetary magnetic fields observed by geocentric
  spacecraft during 1963-1975}.
\newblock {\em Journal of Geophysical Research}, 84(A11):6633--6640.

\bibitem[Cid et~al., 2016]{Cid_2016_ICME_Boundaries}
Cid, C., Palacios, J., Saiz, E., and Guerrero, A. (2016).
\newblock {Redefining the boundaries of interplanetary coronal mass ejections
  from observations at the ecliptic plane}.
\newblock {\em Astrophysical Journal}, 828(1):11.

\bibitem[Cranmer et~al., 2017]{Cranmer_2017_Solar_Wind_Review}
Cranmer, S.~R., Gibson, S.~E., and Riley, P. (2017).
\newblock {Origins of the Ambient Solar Wind: Implications for Space Weather}.
\newblock {\em Space Science Reviews}, 212(3-4):1345--1384.

\bibitem[Gloeckler et~al., 1998]{Gloeckler_1998_SWICS}
Gloeckler, G., Cain, J., Ipavich, F., Tums, E., Bedini, P., Fisk, L.,
  Zurbuchen, T., Bochsler, P., Fischer, J., Wimmer-Schweingruber, R., Geiss,
  J., and Kallenbach, R. (1998).
\newblock Investigation of the composition of solar and interstellar matter
  using solar wind and pickup ion measurements with swics and swims on the ace
  spacecraft.
\newblock {\em Space Science Reviews}, 86(1):497--539.

\bibitem[{Gringauz}, 1961]{Gringauz_1961_Solar_wind}
{Gringauz}, K.~I. (1961).
\newblock {Some Results of Experiments in Interplanetary Space by Means of
  Charged Particle Traps on Soviet Space Probes}.
\newblock In {\em Space Research II}, page 539.

\bibitem[{Gringauz} et~al., 1960]{Gringauz_1960_Solar_Wind}
{Gringauz}, K.~I., {Bezrokikh}, V.~V., {Ozerov}, V.~D., and {Rybchinskii},
  R.~E. (1960).
\newblock {A Study of the Interplanetary Ionized Gas, High-Energy Electrons and
  Corpuscular Radiation from the Sun by Means of the Three-Electrode Trap for
  Charged Particles on the Second Soviet Cosmic Rocket}.
\newblock {\em Soviet Physics Doklady}, 5:361.

\bibitem[{Gringauz} et~al., 1967]{Gringauz_1967_Solar_wind}
{Gringauz}, K.~I., {Bezrukikh}, V.~V., and {Musatov}, L.~S. (1967).
\newblock {Solar-Wind Observations with the Venus 3 Probe}.
\newblock {\em Cosmic Research}, 5:216.

\bibitem[{Heidrich-Meisner} et~al., 2016]{Heidrich_2016_Composition}
{Heidrich-Meisner}, V., {Peleikis}, T., {Kruse}, M., {Berger}, L., and
  {Wimmer-Schweingruber}, R. (2016).
\newblock {Observations of high and low Fe charge states in individual solar
  wind streams with coronal-hole origin}.
\newblock {\em Astronomy \& Astrophysics}, 593:A70.

\bibitem[{Hundhausen}, 1972]{Hundhausen_1972_Solar_Wind_review}
{Hundhausen}, A.~J. (1972).
\newblock {Coronal Expansion and Solar Wind}.
\newblock {\em Physics and Chemistry in Space}, 5.

\bibitem[{Jian} et~al., 2006]{Jian_2006_ICME_Catalog}
{Jian}, L., {Russell}, C.~T., {Luhmann}, J.~G., and {Skoug}, R.~M. (2006).
\newblock {Properties of Interplanetary Coronal Mass Ejections at One AU During
  1995 2004}.
\newblock {\em Solar Physics}, 239(1-2):393--436.

\bibitem[{Jian} et~al., 2011]{Jian_2011_ICME_Catalog}
{Jian}, L.~K., {Russell}, C.~T., and {Luhmann}, J.~G. (2011).
\newblock {Comparing Solar Minimum 23/24 with Historical Solar Wind Records at
  1 AU}.
\newblock {\em Solar Physics}, 274(1-2):321--344.

\bibitem[Larrodera and Cid, 2020]{Larrodera_2020_Bimodal_solar_wind}
Larrodera, C. and Cid, C. (2020).
\newblock Bimodal distribution of the solar wind at 1 au.
\newblock {\em Astronomy \& Astrophysics}, 635:A44.

\bibitem[Lepri, 2004]{Lepri_2004_ICME_Composition}
Lepri, S.~T. (2004).
\newblock {Iron charge state distributions as an indicator of hot ICMEs:
  Possible sources and temporal and spatial variations during solar maximum}.
\newblock {\em Journal of Geophysical Research}, 109(A1).

\bibitem[{Lepri} et~al., 2001]{Lepri_2001_ICME_QFe}
{Lepri}, S.~T., {Zurbuchen}, T.~H., {Fisk}, L.~A., {Richardson}, I.~G., {Cane},
  H.~V., and {Gloeckler}, G. (2001).
\newblock {Iron charge distribution as an identifier of interplanetary coronal
  mass ejections}.
\newblock {\em Journal of Geophysical Research}, 106(A12):29231--29238.

\bibitem[Li et~al., 2016]{Li_2016_Solar_wind_Statistical_Analysis}
Li, K.~J., Zhanng, J., and Feng, W. (2016).
\newblock {A Statistical Analysis of 50 Years of Daily Solar Wind Velocity
  Data}.
\newblock {\em Astrophysical Journal}, 151:128.

\bibitem[{Lionello} et~al., 2005]{2005_Lionello_slow_wind}
{Lionello}, R., {Riley}, P., {Linker}, J.~A., and {Miki{\'c}}, Z. (2005).
\newblock {The Effects of Differential Rotation on the Magnetic Structure of
  the Solar Corona: Magnetohydrodynamic Simulations}.
\newblock {\em Astrophysical Journal}, 625(1):463--473.

\bibitem[{Neugebauer} et~al., 2002]{Neugebauer_2002_Solar_Wind}
{Neugebauer}, M., {Liewer}, P.~C., {Smith}, E.~J., {Skoug}, R.~M., and
  {Zurbuchen}, T.~H. (2002).
\newblock {Sources of the solar wind at solar activity maximum}.
\newblock {\em Journal of Geophysical Research}, 107(A12):1488.

\bibitem[{Neugebauer} and {Snyder}, 1966]{Neugebauer_1966_Solar_Wind_Mariner2}
{Neugebauer}, M. and {Snyder}, C.~W. (1966).
\newblock {Mariner 2 Observations of the Solar Wind, 1, Average Properties}.
\newblock {\em Journal of Geophysical Research}, 71:4469.

\bibitem[{Richardson} and {Cane}, 2005]{Richardson_2005}
{Richardson}, I. and {Cane}, H. (2005).
\newblock {Survey of Interplanetary Coronal Mass Ejections in the Near-Earth
  Solar Wind During 1996 -- 2005}.
\newblock In {\em AGU Spring Meeting Abstracts}, volume 2005, pages SH43A--05.

\bibitem[Richardson and Cane, 1995]{Richardson_1995_CME_Temperature}
Richardson, I.~G. and Cane, H.~V. (1995).
\newblock {Regions of abnormally low proton temperature in the solar wind
  (1965-1991) and their association with ejecta}.
\newblock {\em Journal of Geophysical Research}, 100(A12):23397--23412.

\bibitem[Richardson and Cane,
  2004]{Richardson_2004_ICME_Identification_Composition}
Richardson, I.~G. and Cane, H.~V. (2004).
\newblock {Identification of interplanetary coronal mass ejections at 1 AU
  using multiple solar wind plasma composition anomalies}.
\newblock {\em Journal of Geophysical Research}, 109(A9).

\bibitem[{Rossi}, 1991]{Rossi_1991_Interplanetary_plasma}
{Rossi}, B. (1991).
\newblock {The interplanetary plasma.}
\newblock {\em Annual Review of Astronomy and Astrophysics}, 29:1--8.

\bibitem[{Schwenn}, 2006a]{Schwenn_2006_Solar_wind}
{Schwenn}, R. (2006a).
\newblock {Solar Wind Sources and Their Variations Over the Solar Cycle}.
\newblock {\em Space Science Reviews}, 124:51--76.

\bibitem[{Schwenn}, 2006b]{2006_Schwenn_solar_wind_review}
{Schwenn}, R. (2006b).
\newblock {Space Weather: The Solar Perspective}.
\newblock {\em Living Reviews in Solar Physics}, 3(1):2.

\bibitem[Viall and Borovsky, 2020]{Viall2020}
Viall, N.~M. and Borovsky, J.~E. (2020).
\newblock Nine outstanding questions of solar wind physics.
\newblock {\em Journal of Geophysical Research: Space Physics}.

\bibitem[{Vörös} et~al., 2015]{Voros_2015_Solar_Wind_Statistics}
{Vörös}, Z., {Leitner}, M., {Narita}, Y., {Consolini}, G., {Kov{\'a}cs}, P.,
  {T{\'o}th}, A., and {Lichtenberger}, J. (2015).
\newblock {Probability density functions for the variable solar wind near the
  solar cycle minimum}.
\newblock {\em Journal of Geophysical Research}, 120(8):6152--6166.

\bibitem[{Zhao} et~al., 2009]{Zhao_2009_composition}
{Zhao}, L., {Zurbuchen}, T.~H., and {Fisk}, L.~A. (2009).
\newblock {Global distribution of the solar wind during solar cycle 23: ACE
  observations}.
\newblock {\em Geophysical Research Letters}, 36(14):L14104.

\end{thebibliography}

\end{document}